\documentclass[11pt]{article}
\usepackage{amsmath}
\usepackage{graphicx}
\usepackage{amssymb}
\usepackage{amsfonts}
\usepackage{latexsym}
\def\be{\begin{equation}} \def\ee{\end{equation}}
\def\ba{\begin{eqnarray}} \def\ea{\end{eqnarray}} \def\part{\partial}

\begin{document}

\begin{center}
\begin{flushright}\begin{small}    UFES 2011
\end{small} \end{flushright} \vspace{1.5cm}
\huge{Equivalence of the Hawking temperature in conformal frames} 
\end{center}

\begin{center}
{\small \bf Glauber Tadaiesky Marques $^{(a)}$}\footnote{E-mail address:
gtadaiesky@hotmail.com}\ and
{\small \bf Manuel E. Rodrigues $^{(b)}$}\footnote{E-mail
address: esialg@gmail.com} \vskip 4mm

(a) \ Universidade Federal Rural da Amaz\^{o}nia-Brazil\\
ICIBE - LASIC\\
Av. Presidente Tancredo Neves 2501\\ CEP66077-901 -
Bel\'em/PA, Brazil
\vskip 2mm (b) \ Universidade Federal do Esp\'{\i}rito Santo \\
Centro de Ci\^{e}ncias
Exatas - Departamento de F\'{\i}sica\\
Av. Fernando Ferrari s/n - Campus de Goiabeiras\\ CEP29075-910 -
Vit\'{o}ria/ES, Brazil \vskip 2mm

\end{center}

\begin{center}
                                       Abstract
\end{center}
\hspace{0,6cm} The conformal invariance of the Hawking temperature, conjectured for the asymptotically flat and stationary black holes by Jacobson and Kang, is semiclassically evaluated for a simple particular case of symmetrical spherically and non asymptotically flat black hole. By using the Bogoliubov coefficients, the metric euclideanization, the reflection coefficient and the gravitational anomaly, as methods of calculating the Hawking temperature, we find that it is invariant under a specific conformal transformation of the metric. We discuss briefly the results for each method.    
\vspace{0,5cm}



\section{Introduction}
\hspace{0,6cm} In the Ted Jacobson and Gungwon Kang  paper \cite{jacobson}, it is proved the conformal invariance of the Hawking temperature in five possibilities of geometrics definitions, related to the surface  gravity of black holes. Despite to the fact that this statement has restrictions, as its non validity in the case of conformal continuations in the black hole event horizon \cite{kirill1}, it seems to be quite comprehensive and efficient in the temperature evaluation in conformal frames. In fact, it is shown that it is valid for stationary and asymptotically flat black holes and that the conformal factor is smooth and finite at the event horizon.\par

The context of the above  theorem is purely geometric, so, we want to show its validity, for a particular case of black hole, of this theorem, in a approach close to a semiclassical thermodynamics  phenomenon proposed by Hawking. The Hawking temperature is originated from quantum processes near the event horizon of black holes \cite{hawking1}, so, this phenomenon is not  purely geometric. Through a semiclassical analysis, i.e, quantizing matter fields and letting the gravitational field as classical background field, it can be obtained the temperature of black holes, by various methods well established in the literature.     
\par
In this paper, we propose to make a simple test of the use of conformal invariance of the Hawking temperature, first, calculating the black holes temperature of the Einstein-Maxwell-Dilaton (EMD) theory in the Einstein frame and later in the conformal frame (string frame), and then comparing the results. In a case, a non asymptotically flat black hole, where we can get an exact solution of Klein-Gordon equation, we calculate semiclassically the Hawking temperature, by the methods of surface gravity, the Bogoluibov coefficients, the metric euclideanization, the reflection coefficient of the quantum scalar field and by the gravitational anomaly. This will give us a greater certainty of the conformal invariance of the Hawking temperature, since it is a thermodynamic quantity derived from a semiclassical analysis, and the Jacobson  demonstration is made with geometrical quantities.
\par
This article is organized as follows.  In the second section, we present the EMD theory, the equations of motion and two special classes of solutions, asymptotically flat (AF) and non asymptotically flat (NAF), discussed later. In the third section, we summarize the method for obtaining the Hawking temperature through surface gravity and the calculus for two classes of special solutions. In the fourth section, we define a conformal transformation of the metric. We also present some models of strings theory obtained by fixing the parameter $\lambda$ of the EMD action, and the parameter $\omega$ of the conformal factor. We obtain the solutions of two conformal special classes and their temperatures, through the surface gravity method. In the fifth section we calculate, for the particular case of the NAF black hole, with $\gamma=0$, the Hawking temperature by the semiclassical methods of the Bogoliubov coefficients, subsection 1, of the metric euclideanization, subsection 2, of the reflection coefficient, subsection 3, and of the gravitational anomaly, subsection 4.  In section 6, we present the conclusion and remarks.      


\section{\large The field equations and the black holes solutions}

\hspace{0,6 cm} For understanding very well the origin, the parameters  and structure of the solutions discussed here, we will construct  the technique of obtaining two classes of solutions previously found \cite{gerard1}.

The action of EMD theory is given by: 
\be
S=\int dx^{4}\sqrt{-g}\left[  \mathcal{R}-2\,\eta_{1}
g^{\mu\nu}\nabla_{\mu}\varphi\nabla_{\nu }\varphi+\eta_{2} \,e^{
2\lambda\varphi}F^{\mu\nu}F_{\mu\nu}\right]  \label{action1}\; ,
\end{equation}
where the first term is the usual Einstein-Hilbert gravitational term, while the second and the third are respectively a kinetic term of the scalar field (dilaton or phantom) and a coupling term between the scalar and the Max\-well fields, with a coupling constant $\lambda$ that we assume to be real. The coupling constant $\eta_1$ can take either the value
$\eta_{1}=1$ ({\it dilaton}) or $\eta_{1}=-1$ ({\it anti-dilaton}). The
Maxwell-gravity coupling constant $\eta_{2}$ can take either the value
$\eta_{2}=1$ ({\it Maxwell}) or $\eta_{2}=-1$ ({\it anti-Maxwell}). This action leads to the following field equations:
\begin{eqnarray}
\nabla_\mu\left[  e^{2\lambda\varphi}
F^{\mu\alpha}\right]&=&0\label{fe1}\; ,\\
\Box\varphi &=& - \frac{1}{2}\eta_{1}\eta_{2}\lambda e^{2\lambda\varphi}F^2
\; ,\label{fe2}\\
R_{\mu\nu}&=&2\eta_{1}\nabla_{\mu}\varphi\nabla_{\nu}\varphi+2\eta_{2}\,
e^{2\lambda\varphi}\left(\frac {1}{4}g_{\mu\nu}F^{2}
-F_{\mu}^{\;\;\sigma}F_{\nu\sigma}\right)\label{fe3}\, .
\end{eqnarray}

From now, we will make use of the same procedure as in \cite{gerard1}. Let us write the static and spherically symmetric line element as
\begin{equation}
dS^{2}=e^{2\gamma(u)}dt^{2}-e^{2\alpha(u)}du^{2}-e^{2\beta(u)}d\Omega^{2}
\label{metric}\; .
\end{equation}

The metric function $\alpha$ can be changed  according to the redefinition of the radial coordinate $u$. Then, we consider the harmonic
coordinate condition
\be 
\alpha = 2\beta + \gamma \,.
\ee

We will also assume that the Maxwell field is purely electric (the purely
magnetic case may be obtained by electromagnetic duality transformation $\varphi\rightarrow -\varphi$, $F \to
e^{-2\lambda\varphi}*F$). Integrating (\ref{fe1}), we obtain
\begin{equation}
F ^{10}(u)=q e^ {-2(\lambda\varphi+2\beta+\gamma)} \qquad
(F ^{2}=-2q^{2}e^{-4\beta-4\lambda\varphi}) \label{s1}\; ,
\end{equation}
with $q$ a real integration constant. Substituting  (\ref{s1}) into the equations of motion, we
obtain the equations of second order 
\begin{eqnarray}
\varphi''  &=& -\eta_{1}\eta_{2}\lambda q ^{2}e^{2\omega} \label{e2bis}\; ,\\
\gamma'' &=&\eta_{2}q ^{2}e^{2\omega} \label{e1bis}\; ,\\
\beta'' &=&e^{2J}-\eta_{2}q ^{2}e^{2\omega} \label{e3bis}\; ,
\end{eqnarray}
with
\be
\omega=\gamma-\lambda\varphi\,, \quad J=\gamma+\beta \,,
\ee
and the constraint equation
\be\label{cons}
\beta^{'2} + 2\beta'\gamma' - \eta_1\varphi'^2 = e^{2J} - \eta_2q^2
e^{2\omega}\,.
\ee

By taking linear combinations of the equations (\ref{e2bis})-(\ref{e3bis}),
this system can be integrated and one gets
\begin{eqnarray}
\varphi (u)&=&-\eta_{1}\lambda\gamma (u) +\varphi_{1}u+\varphi_{0}\; ,
\label{e1tertio}\\
\omega'^2 -Qe^{2\omega}&=& a^2\label{e2tertio}\; ,\\
J'^{2}-e^{2J}&=&b^{2}\label{e3tertio}\; ,
\end{eqnarray}
where
\be
\lambda_{\pm}=(1\pm\eta_{1}\lambda^{2})\,, \quad Q=\eta_{2}\lambda_{+}q^2\,,
\ee
and the integration constants $\varphi_{0},\varphi_{1}\in\mathbb{R}$,
$a,b\in\mathbb{C}$.

The general solution of (\ref{e2tertio}) is:
\begin{eqnarray}\label{om}
\omega(u)=\left\{\begin{array}{lr}
-\ln \left| \sqrt{\left| Q\right|} a^{-1}\cosh [a(u-u_{0})]\right| \quad
(a\in \mathbb{R}^{+}\; , \;Q\in\mathbb{R}^{-})\; , \\
a(u-u_{0}) \qquad (a\in\mathbb{R}^{+}\;, \; Q=0 )\; ,\\
-\ln \left| \sqrt{Q} a^{-1}\sinh [a(u-u_{0})]\right| \quad (a\in \mathbb{R}^+
\;,  \;Q\in\mathbb{R}^{+})\; ,\\
-\ln \left| \sqrt{Q} (u-u_{0})\right| \quad (a=0,\; Q\in\mathbb{R}^{+})\;,\\
-\ln \left| \sqrt{Q} \bar{a}^{-1}\sin [\bar{a}(u-u_{0})]\right| \quad (a
=i\bar{a},\; \bar{a},Q\in\mathbb{R}^{+})
\end{array}\right.
\end{eqnarray}
(with $u_0$ a real constant). The general solution of (\ref{e3tertio}) reads:
\begin{equation}\label{J}
J(u)=\left\{\begin{array}{lr}
-\ln \left|  b^{-1}\sinh [b(u-u_{1})]\right| \quad (b\in\mathbb{R}^{+})\;,\\
-\ln \left| u-u_{1}\right| \quad (b=0)\; ,\\
-\ln \left| \bar{b}^{-1}\sin [\bar{b}(u-u_{1})]\right| \quad (b=i\bar{b} ;\;
\bar{b}\in\mathbb{R}^{+})
\end{array}\right.
\end{equation}
($u_1$ real constant). In this way, we have for $\lambda_+\neq 0$, the static and spherically  general solution of
the theory given by the action (\ref{action1}):
\begin{eqnarray}
\label{gs} \left\{\begin{array}{lr}
dS^{2}=e^{2\gamma}dt^{2}-e^{2\alpha}du^{2}-e^{2\beta}d\Omega^{2}\; ,\\
\alpha (u)=2J(u)-\gamma (u)\; ,\\
\beta (u)=J(u)-\gamma (u)\; , \\
\gamma (u)=\lambda _{+}^{-1}(\omega(u)+\lambda\varphi_{1}u+\lambda\varphi_0)
\; ,\\
\varphi (u)=\lambda _{+}^{-1}(-\eta_{1}\lambda\omega(u)+\varphi _{1}u+
\varphi _{0})\; ,\\
F=-q\; e^{2\omega (u)}du\wedge dt\;,
\end{array}\right.
\end{eqnarray}
where, by fixing the spacial infinity at $u_1=0$, we have six integrations constants ($ q, a, b, u_{0}, \varphi _{0}, \varphi_1$), which obey the following  constraint equation (\ref{cons}):
\begin{equation}
\lambda _{+}b^{2}=a^{2}+\eta _{1}\varphi
_{1}^{2}\label{constraint}\; .
\end{equation}

Now, we will choose two particular solutions of the class of that obtained in \cite{gerard1}. According to \cite{gerard1}, if we fix the solutions of $\omega(u)$ and $J(u)$ to be $\sinh$ functions, with the nondegenerate horizon ($m=n=1$) and $u_0>0$, and making the change of radial coordinate 
\begin{equation}\label{30}
u=\frac{1}{(r_{+}-r_{-})}\ln \left( \frac{f_{+}}{f_{-}}\right)\; ,\qquad
f_{\pm}=1- \frac{r_{\pm}}{r}\; ,
\end{equation}
with
\begin{equation}\label{1}
r_\pm = \pm \frac{2a}{1 - e^{\mp 2au_{0}}} \qquad (r_{+}-r_{-} = 2a)\;,
\end{equation}
we obtain the solution 
\begin{eqnarray}
dS^{2}_{AP}&=&f_{+}f_{-}^{\gamma}dt^{2}
-f_{+}^{-1}f_{-}^{-\gamma}dr^{2}
-r^{2}f_{-}^{1-\gamma}d\Omega^{2} \;  ,\label{2} \\
F&=&-\frac{q}{r^2}dr\wedge dt\; ,\;
e^{-2\lambda\varphi}=f_{-}^{1-\gamma}\label{3}
\; ,
\end{eqnarray}
where $\varphi_0=0$, $0<r_{-}<r_{+}$ and $\gamma=\lambda_-/\lambda_+$.
\par
This is an exact solution of a spherically symmetric, asymptotically flat (AF), static and electrically charged black hole, with internal (singular\footnote{In the phantom case the causal structure is discussed in detail in \cite{gerard1}.}) horizon $r_-$ and event horizon $r_+$. The parameters $r_+$ and $r_-$ are related to the physical mass and electric charge for:
\begin{align}
M_{AF} &  =\frac{r_{+}+\gamma r_{-}}{2}\label{4}\; ,\\
q_{AF} &  =\sqrt{\frac{1+\gamma}{2}}\sqrt{\eta_{2}r_{+}r_{-}}\;
.\label{5}
\end{align}

Another solution is obtained when we fix $\omega(u)$ and $J(u)$ as $\sinh$ functions, with the nondegenerate horizon ($m=n=1$) and $u_0=\varphi_0=0$, and making the change of radial coordinate $e^{2bu}=f_+$, we obtain the following solution  
\begin{eqnarray}
dS_{NAF}^{2}=\frac{r^{\gamma}(r-r_+)}{r_{0}^{1+\gamma}}dt^{2}-\frac{r_{0}^{1+\gamma}%
}{r^{\gamma}(r-r_+)}dr^{2}-r_{0}^{1+\gamma}r^{1-\gamma}d\Omega^{2}\, , \label{6}\\
F=-\sqrt{\frac{1+\gamma}{2}}\frac{1}{r_0}dr\wedge dt \;\; ,\;\; e^{2\lambda\varphi}=\left(\frac{r}{r_0}\right)^{1-\gamma}\; .\label{7}
\end{eqnarray}

This is an exact solution of a spherically symmetric, non asymptotically flat (NAF), static and electrically charged black hole, with event horizon $r_+$ ($r_+\geq 0$). The parameters $r_+$ and $r_0$ are related to the physical mass\footnote{We use the quasilocal mass defined in \cite{booth,hawking2}, because it is a non asymptotically flat spacetime.} and the electric charge by:
\begin{align}
M_{NAF} &  =\frac{\left(  1-\gamma\right)  }{4}r_+\label{8}\; ,\\
q_{NAF} &  =r_{0}\sqrt{\frac{\left(  1+\gamma\right)  }{2}}\, .\label{9}
\end{align}
Here, the parameter $r_0$ is related to the electric charge.


\section{\large Hawking temperature by surface gravity}

\hspace{0,6 cm} In this section, we will focus our attention on the geometrical analysis, putting out the gravitational semi-classical effects of the black holes solutions cited in the preciously section, i.e, quantizing others fields called matter fields, and letting the gravitational field as a background. Then, we will make use of the semiclassical thermodynamics of black holes, initiated by Hawking \cite{hawking1} and developed by other authors \cite{davies}. 
\par
There are several ways of obtaining the Hawking temperature, for example, by the Bogoliubov coefficients \cite{ford,glauber}, the metric euclideanization \cite{hawking3}, the energy-momentum tensor \cite{davies}, the reflection coefficient \cite{glauber2,kanti}, the analysis of the gravitational anomaly term \cite{robinson} and the surface gravity of black holes \cite{jacobson}. Until now, these methods have been shown to be equivalent, so we chose to use  in this section the calculation method of the Hawking temperature via the surface gravity for verifying the statement made in \cite{jacobson}.

The surface gravity of a black hole is given by \cite{wald}:
\begin{equation}
\kappa =\left[  \frac{g_{00}^{\prime}}{2\sqrt{-g_{00}g_{11}}}\right]_{r\;=\;r_{H}}\, ,\label{10}
\end{equation}
where $r_{H}\;$ is the event horizon radius, and the Hawking temperature is related with the surface  gravity through the relationship \cite{jacobson,hawking1}
\begin{equation}
T=\frac{\kappa}{2\pi}\, .\label{11}
\end{equation}

Then, for the AF black hole case (\ref{2}), we get the surface gravity (\ref{10}) as 
\begin{equation}
 \kappa_{AF} =\frac{\left(  r_{+}-r_{-}\right)  ^{\gamma}}{2r_{+}^{1+\gamma}}\, ,\label{12}
\end{equation}
and the Hawking temperature (\ref{11}) in this case is :
\begin{equation}
T_{AF}=\frac{\left(  r_{+}-r_{-}\right)  ^{\gamma}}{4\pi r_{+}^{1+\gamma}}\, .\label{13}
\end{equation}

Using the metric of NAF black hole (\ref{6}) in (\ref{10}), we obtain:
\begin{align}
\kappa_{NAF} =\frac{r_{+}^{\gamma}}{2r_{0}^{1+\gamma}}\, ,\label{14}
\end{align}
substituting (\ref{14}) in (\ref{11}), we get the Hawking temperature for the NAF black hole (\ref{6}):
\begin{equation}
T_{NAF}=\frac{\kappa_{NAF}}{2\pi}=\frac{r_{+}^{\gamma}}{4\pi r_{0}^{1+\gamma}}\, .
\label{15}
\end{equation}

Defining the black holes entropy as $S=\frac{1}{4}A=\pi r_{0}^{1+\gamma}r^{1-\gamma}$, where $A(r=r_+)=\int\int\sqrt{g_{22}g_{33}}d\theta d\phi$ is the surface of event horizon of the black hole, and through (\ref{7}) we obtain the scalar electric potential as $A_{0}(r_{+})=-\sqrt{(1+\gamma)/2}\,(r_{+}/r_{0})$. The first law of the thermodynamics of charged black holes, for the NAF case (\ref{6}), is given \cite{carter}
\begin{equation}
dM_{NAF}=T_{NAF}dS+A_{0}(r_{+})dq_{NAF}\; ,
\end{equation}
where we have to fix a value for the electric charge $q_{NAF}$ ($dq_{NAF}=0$), or fixing a gauge for the electrical potetial \cite{cedric},  to satisfy the first law.

\section{\large Hawking temperature for the String models at low energies}

\hspace{0,6 cm} In this section, we will evaluate the Hawking temperature for some strings  models at low energies coupled only with the electromagnetic field. The semiclassical analysis for these strings models is possible because it is done on an energy scale that is valid for both  the semiclassical analysis and the string theory in four dimensions. We will pass from the EMD theory to the String Theory (ST) at low energies by a conformal transformation, where in the literature the terms {\it Einstein frame} is used for EMD theory, while {\it Strings frame} is used for the ST models. Also, it will be clear later why  we will calculate the Hawking temperature for the ST models; the parameter $\lambda$ that we introduced in the action (\ref{action1}) can vary on any real value, so getting, for a fix chosen value of $\lambda$, the possible ST models.   
\par
We are now interested in obtaining the strings models to black holes for both the AF (\ref{2}) and the NAF (\ref{6}).  To do this we will firstly analyse the action (\ref{action1}), through a generalized conformal transformation. \par

A conformal transformation leads  a manifold ``$M$", with metric ``$g_{\mu\nu}$" , to another manifold ``$\widehat{M}$", with metric ``$\widehat{g}_{\mu\nu}$", preserving angles and ratios between geometric objects. The relationship between the metrics is given by the function $\Omega (x^{\alpha})$ which is differentiable and strictly positive:
\begin{equation}
\widehat{g}_{\mu\nu}=\Omega^{2}g_{\mu\nu} \, ,\label{16}
\end{equation}
then, in $4$ dimensions, we get
\begin{align}
g^{\mu\nu} & =\Omega^{2}\widehat{g}^{\mu\nu}\; ,\; \sqrt{-g} =\Omega^{-4}\sqrt{-\widehat{g}}\label{17}\; ,\\
\mathcal{R} & =\Omega^{2}\widehat{\mathcal{R}}+6\Omega\,  \widehat{g}^{\mu\nu}\widehat{\nabla}_{\mu}
\widehat{\nabla}_{\nu}\Omega\; .\label{18}
\end{align}

In order to obtain the strings models, we choose a general conformal factor
\begin{equation}
\Omega=e^{ -\omega\varphi}\; ,\label{19}
\end{equation}
where $\omega$ is a new real parameter whose the introduction will enable us to obtain the Strings models. So, replacing (\ref{19}) in (\ref{17}) and (\ref{18}), we obtain:
\begin{align}
g^{\mu\nu} & =e^{-2\omega\varphi}\widehat{g}^{\mu\nu}\; , \;\sqrt{-g} =e^{4\omega\varphi} \sqrt{-\widehat{g}}\label{20}\\
\mathcal{R} & =e^{-4\omega\varphi}\left[ \widehat{\mathcal{R}}+6\omega^{2}\widehat{g}^{\mu\nu}\widehat{\nabla}_{\mu}\varphi\widehat{\nabla}_{\nu}\varphi\right] \, .\label{21}
\end{align}

Making a conformal transformation on the metric, through the expressions (\ref{20}) and (\ref{21}), the action (\ref{action1}) becomes
\begin{align}
S =\int d^{4}x\sqrt{-\widehat{g}}\;e^{2\omega\varphi}\left\{  \widehat{\mathcal{R}}+2\left( 3\omega^{2}-\eta_{1}\right)  \widehat{g}^{\mu\nu}\widehat{\nabla}_{\mu}\varphi\widehat{\nabla}_{\nu}\varphi+\eta_{2}e^{2\varphi\left(
\lambda-\omega\right)  }\widehat{F}^{2}\right\}, \label{27}
\end{align}
where
\begin{equation}
\widehat{g}_{\mu\nu} =\Omega^{2}g_{\mu\nu}=e^{-2\omega\varphi}g_{\mu\nu
}=\left[h(r)\right]^{\frac{\omega
}{\lambda}\left(  1-\gamma\right)  }g_{\mu\nu}\label{26}\; ,\\
\end{equation}
and $h(r)=f_{-}$, for the AF black hole (\ref{2}), and $h(r)=r/r_{0}$ for the NAF black hole (\ref{6}). The action (\ref{27}), which is in the {\it Strings frame}, can be classified into the three following types:

\begin{enumerate}

\item Type I Strings

As we need the usual Strings models\footnote{We could formulate the Strings models with phantom contribution \cite{hull}.},  making use of the dilaton case in which $\omega=\eta_{1,2}=1,\, \lambda=\frac{1}{2}$, the action (\ref{action1}) reads:
\begin{equation}
S_{I}=\int d^{4}x\sqrt{-\widehat{g}}\left\{e^{-2\varphi}\left[  \widehat{\mathcal{R}}+4\widehat{g}^{\mu\nu}\widehat{\nabla}_{\mu}\varphi
\widehat{\nabla}_{\nu}\varphi\right]-e^{-\varphi}
\widehat{F}^{2}\right\}, \label{28}
\end{equation}
which is just the type I Strings action in $4$ dimensions.
\item Type IIA Strings

For the dilaton case, we chose the parameters $\omega=\eta_{1,2}=1,\, \lambda=0$, and the action (\ref{action1}) becomes:
\begin{equation}
S_{IIA}=\int d^{4}x\sqrt{-\widehat{g}}\left\{ e^{-2\varphi}\left[\widehat{\mathcal{R}}+4\widehat{g}^{\mu\nu}\widehat{\nabla}_{\mu}\varphi
\widehat{\nabla}_{\nu}\varphi\right]-\widehat{F}^{2}\right\}. \label{29}
\end{equation}
This is known as the type IIA Strings action in $4$ dimensions.
\item Heterotic Strings

For the dilaton case, we chose the parameters $\pm\omega=\eta_{1,2}=\mp\lambda=1$, and the action  (\ref{action1}) behaves as:
\begin{equation}
S_{H}=\int d^{4}x\sqrt{-\widehat{g}}\;e^{-2\varphi}\left\{\widehat{\mathcal{R}}+4\widehat{g}^{\mu\nu}\widehat{\nabla}_{\mu}\varphi
\widehat{\nabla}_{\nu}\varphi-\widehat{F}^{2}\right\}, \label{30}
\end{equation}
which is just the Heterotic String action in $4$ dimension. 
\end{enumerate}

Now we will get the line elements (\ref{2}) and (\ref{6}) in  Strings frame, using the conformal transformation (\ref{26}). The line element (\ref{2}), transformed conformally by (\ref{26}), with $h(r)=f_{-}$, becomes
\begin{align}
d\widehat{S}_{AF}^{2} & = f_{+} f_{-}^{\frac{\omega}{\lambda}+\gamma(1-\frac{\omega}{\lambda})} dt^{2}-f_{+}^{-1} f_{-}^{\frac{\omega}{\lambda}-\gamma(1+\frac{\omega}{\lambda})} dr^{2} -r^{2}f_{-}^{(1-\gamma)(1+\frac{\omega}{\lambda})}d\Omega^{2} \label{31}\, .
\end{align}

The parameters $r_{+}$ and $\, r_{-}$ are related with the physical mass and charge by:
\begin{align}
\widehat{M}_{AF} & =\frac{1}{2}\left\{ r_{+}+r_{-}\left[ \frac{\omega}{\lambda}+\gamma\left(1-\frac{\omega}{\lambda}\right)\right]\right\}\label{32}\; ,\\
\widehat{q}_{AF} & =\sqrt{\frac{1+\gamma}{2}}\sqrt{\eta_{2}r_{+}r_{-}}\label{33}\; .
\end{align}

The line element (\ref{6}), conformally transformed by (\ref{26}), becomes
\begin{align}\label{34}
d\widehat{S}_{NAF}^{2} & =\frac{(r-r_+)r^{\frac{\omega}{\lambda}+\gamma\left( 1-\frac{\omega}{\lambda}\right)}}{r_{0}^{\left(1+\frac{\omega}{\lambda}\right)+\gamma\left( 1-\frac{\omega}{\lambda}\right)}}dt^{2}-\frac{r^{\frac{\omega}{\lambda}-\gamma\left( 1+\frac{\omega}{\lambda}\right)}r_{0}^{\left(1-\frac{\omega}{\lambda}\right)+\gamma\left( 1+\frac{\omega}{\lambda}\right)}}{(r-r_+)}dr^{2}\nonumber\\
&-\frac{r^{(1-\gamma)\left( 1+\frac{\omega}{\lambda}\right)}}{r_{0}^{\left[ (1-\gamma)\left( 1+\frac{\omega}{\lambda}\right)-2\right]}}d\Omega^{2}  \;.
\end{align}

The parameters $r_{+}$ and $r_{0}$ are related with the physical parameters mass\footnote{The quasilocal mass.} and charge by:
\begin{align}
\widehat{M}_{NAF} & =\left( \frac{1-\gamma}{4}\right)r_{+}\label{35}\; ,\\
\widehat{q}_{NAF} & = r_{0}\sqrt{\frac{\left(  1+\gamma\right)  }{2}}\label{36}\; .
\end{align}

Now we can calculate the surface gravity of the AF (\ref{31}) and NAF (\ref{34}) black holes. The surface gravity (\ref{10}) of the conformal black hole (\ref{31}) is given by:
\begin{align}
\widehat{\kappa}_{AF} &  =\left[  \frac{\widehat{g}_{00}^{\prime}}%
{2\sqrt{-\widehat{g}_{00}\widehat{g}_{11}}}\right]  _{r\;=\;r_{H}%
}\nonumber\\
\widehat{\kappa}_{AF}  &  =\kappa_{AF}+\left[  \frac{\Omega^{\prime}}{\Omega}%
\sqrt{-\frac{g_{00}}{g_{11}}}\right]  _{r=r_{H}}. \label{37}
\end{align}

Then, the simple calculation of the second term on the right side of (\ref{37}), determines the surface gravity of the conformal black hole (\ref{31}).  Using (\ref{2}) and (\ref{26}), we have:
\begin{equation}
\left[  \frac{\Omega^{\prime}}{\Omega}\sqrt{-\frac{g_{00}}{g_{11}}}\right]_{r=r_{H}}=0\label{38},
\end{equation}
thus, from (\ref{12}), (\ref{37}) and (\ref{38}), the Hawking temperature of the conformal black hole (\ref{31}) reads:
\begin{equation}
\widehat{T}_{AF}=\frac{\widehat{\kappa}_{AF}}{2\pi}=T_{AF}=\frac{\left(
r_{+}-r_{-}\right)  ^{\gamma}}{4\pi r_{+}^{1+\gamma}} \, .\label{39}
\end{equation}

We initially calculated the Hawking temperature for the black hole (\ref{2}), which is found in {\it Einstein frame}, resulting into (\ref{13}). Now, we calculated the Hawking temperature for the black hole (\ref{31}), which is in the Strings frame, resulting into (\ref{39}). This shows us a very useful property, that is the conformal invariance of the Hawking temperature of black holes generated by (\ref{2}) and (\ref{31}).       
\par
Now, we calculate the surface gravity (\ref{10}) for the black hole (\ref{34}), and get:
\begin{equation}
 \widehat{\kappa}_{NAF} =\kappa_{NAF}=\frac{r_{+}^{\gamma}}{2r_{0}^{1+\gamma}}\, ,\label{40}
\end{equation}
and the Hawking temperature (\ref{11}), in this case, becomes
\begin{equation}
\widehat{T}_{NAF}=T_{NAF}=\frac{r_{+}^{\gamma}}{4\pi r_{0}^{1+\gamma}}\, .\label{41}
\end{equation}

Thus, we obtain the conformal invariance of the Hawking temperature of the black holes (\ref{6}) and (\ref{34}). The first law of the thermodynamics is not satisfied here because the entropy of the conformal case is not conformal invariant, despite to the existence of methods that make it invariant \cite{vanzo2}.

\section{\large The semiclassical computation of the Hawking temperature}

\hspace{0,6cm} In this section, we will calculate the value of the Hawking temperature, for the black holes (\ref{6}) and (\ref{34}), in the particular case $\lambda=\eta_1=+1$ ($\gamma=0$), through of the methods of Bogoliubov coefficients, the metric euclideanization, the reflection coefficient and gravitational anomaly. 

\subsection{\large Hawking temperature from the Bogoliubov coefficients}

\hspace{0,6cm} The calculation of Hawking temperature using the Bogoliubov coefficients will be realised following the references \cite{ford,glauber}. The unique case for which we will analyse the value of the Hawking temperature of the various methods is the ANF black hole (\ref {6}), for $\gamma=0$ ($\lambda=\eta_1=\eta_2=+1$). But as the line element (\ref{6}) is a particular case, for $\omega=0$ in (\ref{34}), we will make all calculations with the line element (\ref{34}), for $\gamma=0$.
\par
First, we will make the glue of the metric by the collapse of a spherical fine shell, analysing only the second collage. The modes in ($v$), arising from the infinite past $\cal{I}^{-}$, of a quantized scalar field, enter in the spherical shell and pass through a flat region \footnote{We will obtain the same result if we consider the vacuum of the  EMD theory of the black hole (\ref{6}).} of space-time, and then leave as out modes ($u(v)$), escaping to the future infinite $\cal{I}^{+}$.
\par
By doing so the second collage of the metric, of the asymptotically flat regions with the non-asymptotically flat ones, we obtain
\begin{eqnarray}
dS_{Minkowski}^{2} &=& d\widehat{S}_{ANP}^{2}\nonumber\\
dT^2-dR^2-R^2d\Omega^2 &=& \frac{R^{\omega}(R-r_{+})}{r_{0}^{1+\omega}}dt^2-\frac{r_{0}^{1-\omega}R^{\omega}}{(R-r_{+})}dR^2-r_{0}^{1-\omega}R^{1+\omega}d\Omega^2\nonumber \; .
\end{eqnarray}       
For a spacetime without rotation, $d\Omega/dT=0$, and the radial coordinate approximated near the horizon, $R\cong r_{+}+c(T_{0}-T)$, one gets 
\begin{equation}
t=\pm r_{0}\ln (T_{0}-T)\label{tm}\; .
\end{equation}

The coordinate  $r^{*}=\sqrt{-\widehat{g}_{11}/\widehat{g}_{00}}=r_{0}\ln (T_{0}-T)$, yields the retarded temporal coordinate $u(T)=t-r^{*}=-2r_{0}\ln (T_{0}-T)$, for the choice of the sign  minus in (\ref{tm}). As in the interior of the shell $U=V=v+c_{0}=T-R$, then $T(v)=(v+c_{1})/(1+c)$,  with $c_{1}=c_{0}+cT_{0}+r_{+}$. Finally  we get
\begin{equation}
u(v)=-2r_{0}\ln \left( \frac{v_{0}-v}{1+c}\right)\label{u}\; ,
\end{equation}
where $v_{0}=T_{0}-c_{0}-r_{+}$. As we can write  $u(v)=-\widehat{\kappa}^{\,-1}\ln \left(\frac{v_{0}-v}{1+c}\right)$, we recognize that the surface gravity of the black hole is
$\widehat{\kappa} =(1/2r_{0})$. Continuing the calculation as in \cite{ford}, we will get the Hawking temperature given by
\begin{equation}
\widehat{T}_{H}=\frac{\widehat{\kappa}}{2\pi}=\frac{1}{4\pi r_{0}}\label{th1}\; .
\end{equation}
As this result is independent on the parameter $\omega$ of the conformal factor, we have that for $\omega=0$, the line element becomes (\ref{6}), for $\gamma=0$, and Hawking temperature is exactly the same as (\ref{th1}). We then show the conformal invariance of the Hawking temperature which in this particular case is calculated semi-classically.
\par
The main reason for the conformal invariance of the temperature in the calculation presented here is that the collage of the metric gives us a time $t(T)$  which is independent on the conformal factor, because it is eliminated in the near horizon limit. Also, the tortuous coordinate
 $r^{*}=\sqrt{-\widehat{g}_{11}/\widehat{g}_{00}}= \sqrt{-g_{11}/g_{00}}$, is clearly invariant, thus providing a retarded temporal coordinate $u(v)$, which is independent from the conformal factor. This result is not clear from the beginning of the evaluation of the temperature through this method.
\subsection{\large Calculating the Hawking temperature through metric euclidianization}

\hspace{0,6cm} The calculation of the Hawking temperature via the metric euclidianization method \cite{hawking2}, basically consists of making an analytical continuation of the metric, transforming the temporal coordinate into a pure imaginary one, belonging to the set of complex numbers. So, the space-time without the angular extent, represents the geometry of a conical surface, with the famous singularity at the nipple. To avoid this singularity, it is necessary to impose the new coordinate related to the time transformation to be periodic, thus resulting into a constraint between the surface gravity and the Hawking temperature.
\par
Making the analytical continuation of the metric (\ref{34}), for the case
 $\lambda=\eta_1=\eta_2=+1$, by the transformation $dt=id\tau/\alpha$, we get 
\begin{eqnarray}
dS_{ANP(E)}^2 &=& -\left[ \frac{r^{\omega}(r-r_{+})}{r_{0}^{1+\omega}}\frac{d\tau^2}{\alpha^2}+\frac{r_{0}^{1-\omega}r^{\omega}}{(r-r_{+})}dr^2\right]\label{mec1}\; ,\\
&=& -\Omega(\rho)\left[ \rho^2d\tau^2+d\rho^2\right]\label{mec2}\; ,
\end{eqnarray}
ignoring the angular part of the metric. Comparing the first term of both the line elements, we have $\Omega=r^{\omega}(r-r_{+})/\alpha^{2}r_{0}^{1+\omega}\rho^2$.  Doing the same thing for the second term of the  line elements, we have $\rho=e^{\pm\alpha  r ^{*}}$, with $r^{*}=\sqrt{-\widehat{g}_{11}/\widehat{g}_{00}} = r_{0}\ln(r-​​r_{+})$. Remembering that we can describe $r^{*}$  in terms of the surface gravity of the black hole as $r^{*}= (1/2\widehat{\kappa})\ln(r-​​r_{+})$ in this case, we have $\widehat{\kappa} = (1/2r_{0})$, and to avoid the canonical singularity, the conformal factor must be finite at the event horizon. Choosing the sign plus  for $\rho$, we obtain
\begin{equation}
\Omega(r)=\frac{r^{\omega}(r-r_{+})}{\alpha^2r_{0}^{1+\omega}(r-r_{+})^{(\alpha /\widehat{\kappa})}}\; .\label{fce}
\end{equation}

The canonical singularity is avoided only if $\alpha=\widehat{\kappa}=(1/2r_{0})$. The Hawking temperature, calculated by the metric euclidianization method is given by
\begin{equation}
T_{H}=\frac{\alpha}{2\pi}=\frac{1}{4\pi r_{0}}\label{the}\; .
\end{equation}
One again, the main feature of the conformal invariance of Hawking temperature, by the metric euclidianization method is given by the conformal invariance of the tortuous coordinate $ r^{*}$, since the conformal factor always depends on the inverse of the new radial coordinate  $\rho$ by the expression $\Omega=\widehat{g}_{00}/\alpha^2\rho^2$. This tells us that when the conformal factor (\ref{26}) does not depend on powers of $(r-​​r_{+})$, the new conformal factor in (\ref{mec2}) is finite when $\alpha=\widehat{\kappa}=\kappa$.
\subsection{\large Hawking temperature through the reflection coefficient}
\hspace{0,6cm} 
In this subsection, we will obtain the temperature of the ANF black hole  (\ref{34}), for the case $\lambda=\eta_1=\eta_2=+1$, by the reflection coefficient method of the quantum field modes \cite{kanti}. This method can be used more effectively when we can solve exactly the Klein-Gordon equation for the scalar field. After, it is taken the infinite space and the near event horizon limits, in the Klein-Gordon equation, obtaining approximate solutions. Comparing the constants of the exact and approximate solutions in these limits, one can calculate the transmission and reflection coefficients, through the field flow. Thus, one can calculate the Hawking temperature in the limit of high frequencies, by the reflection coefficient.
\par  
The Klein-Gordon equation for a non massive quantum scalar field is given by
\begin{align}
\square \varphi =0 \label{42}\; .
\end{align}
 
Considering the dependence of the field as
\begin{equation}
\varphi (t,r,\theta,\phi)=Y(\theta,\phi)R(r)\,e^{-i\omega_{1} t}\; ,
\end{equation}
the equation (\ref{42}), for the metric (\ref{34}), becomes 
\begin{align}
&r(r-r_{+})\partial^{2}_{r}R+\left\{(r-r_{+})\left[1+\frac{\omega}{\lambda}(1-\gamma)\right]+r\right\}\partial_{r}R\nonumber\\
&+\left[\frac{\bar{\omega}^{2}r^{1-2\gamma}}{(r-r_{+})}-l(l+1)\right]R=0\; .\label {44}
\end{align}
where $\bar{\omega}=r_{0}\omega_{1}$. Making the change of coordinates  $y=(r_{+}-r)/r_{+}$ and $R=y^{i\bar\omega}f(y)$, for the case $\gamma=0$, we get:
\begin{align}
y(1-y)f^{\prime\prime}+[1+2i\bar{\omega}-(2i\bar{\omega}+1+a)y]f^{\prime}+[l(l+1)+i\bar{\omega}a]f=0\label {45}
\end{align} 
with $a=1+\omega$. The general solution is  
\begin{align}
R(r)=&C_{1}\left(\frac{r_{+}-r}{r_{+}}\right)^{i\bar{\omega}}F\left(d_{+};e_{+};c_{+};\frac{r_{+}-r}{r_{+}}\right)+C_{2}\left(\frac{r_{+}-r}{r_{+}}\right)^{-i\bar{\omega}}\times\nonumber\\
&\times F\left(d_{-};e_{-};c_{-};\frac{r_{+}-r}{r_{+}}\right)\label {46},
\end{align}
where $d_{\pm}=\pm i\bar{\omega}+a/2- i\alpha\; , e_{\pm}=\pm i\bar{\omega}+a/2+ i\alpha\; , c_{\pm}=1\pm2i\bar{\omega}\; , \alpha=\sqrt{\bar{\omega}^{2}+l(l+1)-\left(a/2\right)^{2}}$, $C_1,C_2\in \mathbb{C}$ and the function $F\left(d_{+};e_{+};c_{+};y\right)$ is the hypergeometric function.
\par
We will now study the asymptotic limits of the solution, with the aim of determining the integration constants and the transmission and reflection coefficients. Defining the flow of the scalar field as
\begin{align}
\mathcal{F}^{\mu}=\frac{2\pi}{i}\sqrt{-g}g^{\mu\nu}[R^{*}(r)\partial_{\nu}R(r)-R^{*}(r)\partial_{\nu}R(r)]\label{55}\; ,
\end{align}
and the transmission and reflection coefficients as
\begin{align}
\mathcal{T}=\left| \frac{\mathcal{F}^{(IN)r}_{NEAR}(r\approx r_{+})}{\mathcal{F}^{(IN)r}_{FAR}(r\approx +\infty)}\right|\;\; ,\;\; \mathcal{R}=\left| \frac{\mathcal{F}^{(OUT)r}_{FAR}(r\approx +\infty)}{\mathcal{F}^{(IN)r}_{FAR}(r\approx +\infty)}\right|\label{56}\; ,
\end{align}
we make the asymptotic analysis, using the properties of the hypergeometric function, and comparing the two asymptotic solutions  with the exact solution limits, we get
\begin{align}
\mathcal{T}=\frac{ 2\sinh (2\pi\alpha )\sinh (2\pi\bar{\omega })}{\cosh [2\pi (\bar{\omega}+\alpha )]-\cos (a\pi)}\label{61}\; ,\\
\mathcal{R}=\frac{\cosh [2\pi (\bar{\omega }-\alpha )]-\cos (a\pi)}{\cosh [2\pi (\bar{\omega }+\alpha )]-\cos (a\pi )}\label{62}\; .
\end{align}
\par
The normalization condition $\mathcal{T}+\mathcal{R}=1$ is satisfied. Now we calculate the Hawking temperature through the expressions  $\widehat{T}=-(\omega_1/\ln \mathcal{R})$ and $\widehat{T}_{H}=\lim_{\omega_{1}>>1}\widehat{T} $. Remembering that  $\bar{\omega}=r_{0}\omega_1$, we obtain
\begin{align}
\widehat{T}&=\frac{\omega_1}{\ln \left\{ \frac{\cosh [2\pi (r_{0}\omega_1+\alpha )]-\cos (a\pi)}{\cosh [2\pi (r_{0}\omega_1-\alpha )]-\cos (a\pi )}\right\}}\;,\\
\widehat{T}_{H}&\approx\frac{\omega_1}{4\pi r_{0}\omega_1+\ln [1-\cos (1+\omega)\pi]}\approx \frac{1}{4\pi r_{0}}\; ,\label{thcr}
\end{align}
which agrees with the temperature previously obtained in the Einstein frame \cite{glauber2}.

By this method of obtaining the Hawking temperature, we see clearly, by (\ref{thcr}), that the value of the temperature for this particular case, is a conformal invariant. The main reason in this case, of the conformal invariance, is not clear as in the previous methods. But we can say that the limit of high frequencies $\omega_1$, of the modes of the quantum scalar field, just ignores the contribution caused by the conformal factor.
\par
\subsection{Hawking temperature from gravitational anomaly}

\hspace{0,6cm} A new method for obtaining the Hawking temperature is formulated by Robinson and Wilczek \cite{robinson}. Such methods based on the calculation of the gravitational anomaly (and gauge anomaly for charged black holes) of the energy-momentum tensor, which depends on its flow, because of the modes  emerging from the chiral field of quantum theory \cite{bertlmann}. As this anomaly appears only near the event horizon of the black hole, one has to evaluate it in two regions: $r_{+}+\delta <r$, away from the horizon, and the region $r_{+}\leq r \leq r_{+}+\delta$ near the horizon. The cancellation of gauge and gravitational anomalies depends on the term proportional to the square of the Hawking temperature. Thus, it is possible to cancel anomalies and to obtain Hawking temperature value at the same time. The formulation of the anomaly cancellation for the action of the EMD theory is done extensively in \cite{wu}.
\par
The quantum vacuum used for the expected value of the energy-mo\-men\-tum tensor is that of Unruh \cite{unruh,rabin}. In the region $r_{+}+\delta <r$, the four-momentum conservation \footnote{Even the classical energy-momentum tensor is not conserved in the presence of a four-current.} and that of the four-current obey the classical equations of conservation
\begin{align}
\nabla_{\mu}T^{\mu}_{\nu}=F_{\mu\nu}J^{\mu}\; ,\;\nabla_{\mu}J^{\mu}=0\; .\label{ec}
\end{align}     
\par

In the region near the horizon $r_{+}\leq r\leq r_{+}+\delta$, there is gravitational and gauge anomalies. Thus, the classical equations become anomalous and obey the two-dimensional covariant expressions \cite{rabin,gangopadhyay}
\begin{align}
\nabla_{\mu}\tilde{J}^{\mu} & =\frac{q^2}{2\pi\sqrt{-\widehat{g}}}\partial_{r}A_{t}(r)\label{ac}\; ,\\
\nabla_{\mu}\tilde{T}^{\mu}_{\nu} & =F_{\mu\nu}\tilde{J}^{\mu}+\frac{1}{96\pi\sqrt{-\widehat{g}}}\epsilon_{\nu\mu}\partial^{\mu}R\; ,\label{at}
\end{align}  
where $R$ is the curvature scalar  and $\epsilon_{10}=+1$. Integrating the equations  (\ref{ec})-(\ref{at}) and combining the results, one gets 
\begin{align}
\sqrt{-\widehat{g}}\tilde{J}^{r}(r) & =a_{H}+\frac{q^2}{2\pi}A^2_{t}(r)-\frac{q^2}{4\pi}A^2_{t}(r_{+})\label{j}\; ,
\end{align}
\begin{align}
\sqrt{-\widehat{g}}\;\tilde{T}^{r}_{t}(r) & =b_{H}+\frac{q^2}{4\pi}\left[A^2_{t}(r)-A^2_{t}(r_{+})\right]-\frac{q^2}{2\pi}A_{t}(r_{+})\left[A_{t}(r)-A_{t}(r_{+})\right]\nonumber\\
& +\frac{1}{192\pi}\Bigg\{ 2\left(-\widehat{g}^{\;-1}_{11}\right)\partial_{r}^{2}\left(\widehat{g}_{00}\right)+\partial_{r}\left(\widehat{g}_{00}\right)\partial_{r}\left(-\widehat{g}^{\;-1}_{11}\right)\nonumber\\
& -2\left(-\widehat{g}_{00}\widehat{g}_{11}\right)^{-1}\left[\partial_{r}\left(\widehat{g}_{00}\right)\right]^{2}\Bigg\}+\frac{1}{192\pi}\Bigg\{ \left(-\widehat{g}^{\;-1}_{11}\right)\partial_{r}^{2}\left(\widehat{g}_{00}\right)\nonumber\\
&+\partial_{r}\left(\widehat{g}_{00}\right)\partial_{r}\left(-\widehat{g}^{\;-1}_{11}\right)\Bigg\}_{r=r_{+}}\; \label{t},
\end{align}
where $a_{H}$ and  $b_{H}$  are integration constants.\par
Using the property that the components of the current and that of the covariant energy-momentum tensor cancel out on the horizon, ie, the anomaly is cancelled in a gauge transformation for the current and a coordinate transformation for the energy-mo\-men\-tum tensor, we obtain the constants $a_{H}=-(q^2/4\pi)A^2_{t}(r_{+})$ and $b_{H}=(1/96\pi)\left[\partial_{r}\left(\widehat{g}_{00}\right)\partial_{r}\left(-\widehat{g}_{11}^{\;-1}\right)\right]_{r=r_{+}}$. Substituting these constants into (\ref{j}) and (\ref{t}), and making the asymptotic limit at the infinite space, we get
\begin{align}
\sqrt{-\widehat{g}}\tilde{J}^{r}(r\rightarrow\infty) & =-\frac{q^2}{2\pi}A^2_{t}(r_{+})\label{fj}\; ,\\
\sqrt{-\widehat{g}}\;\tilde{T}^{r}_{t}(r\rightarrow\infty) & =\frac{q^2}{4\pi}A^2_{t}(r_{+})+\frac{\pi}{12}\widehat{T}_{H}^2\; \label{ft},
\end{align}
where the Hawking temperature is given by 
\begin{equation}
\widehat{T}_{H}=\frac{1}{4\pi}\sqrt{\left[\partial_{r}\left(\widehat{g}_{00}\right)\partial_{r}\left(-\widehat{g}_{11}^{\;-1}\right)\right]_{r=r_{+}}}\label{tha}
\;.\end{equation}
In the case of ANF black hole (\ref{34}), for $\lambda=\eta_{1,2}=+1$, with the aid of (\ref{7}), the expressions (\ref{fj}) and (\ref{ft}) become 
\begin{align}
\sqrt{-\widehat{g}}\tilde{J}^{r}(r\rightarrow\infty) & =-\frac{q^2}{2\sqrt{2}\pi}\frac{r_{+}}{r_{0}}\label{fj1}\; ,\\
\sqrt{-\widehat{g}}\;\tilde{T}^{r}_{t}(r\rightarrow\infty) & =\frac{1}{8\pi}\left(\frac{qr_{+}}{r_{0}}\right)^2+\frac{\pi}{12}\left(\frac{1}{4\pi r_{0}}\right)^2\; \label{ft1}.
\end{align} 

Therefore, the Hawking temperature calculated by the gravitational a\-no\-ma\-ly method, for the case of the conformal black hole in (\ref{34}) with $\gamma=0$, is given by $\widehat{T}_{H}=(1/4\pi r_{0})$. As this result does not depend on the parameter $\omega$ of the conformal factor, the temperature calculation for the case of Einstein frame (\ref{6}), presents the same value as in the conformal case. In the case of the Einstein frame, we have $g_{00}=-g_{11}^{-1}$, and so, the expression obtained by the gravitational anomaly is equal to (\ref{10}). The expression (\ref{tha}) gives us again an invariant conformal value, taking into account that the conformal factor $\Omega(r_{+})$ is finite, $g_{00}(r_{+})=0$ and $\partial_{r}(-g_{00}g_{11}^{-1})=0$ for $g_{00}=-g_{11}^{-1}$. It seems that the main feature of conformal invariance of the Hawking temperature, calculated by this method, is the invariance of the second term of the flow of energy-momentum tensor (\ref{ft}), coming from the cancellation of the  covariant gravitational anomaly. This fact seems to be related to the invariance of the flow as the energy-momentum tensor in the Hawking effect, as shown in \cite{jacobson}.

\section{\large Conclusion}


\hspace{0,6cm}
We calculted geometrically the Hawking temperature by the surface gravity method in the sections $3$ and $4$. The  section $3$ is devoted to the two classes of black hole solutions (\ref{2})  and (\ref{6}), in the Eisntein frame, while in the section $4$, (\ref{31}) and (\ref{34}) black hole solutions  are treated in the Strings frame. The result is not  surprising because Jacobson and Kang had already demonstrated that the Hawking temperature, calculated by the surface gravity method, is a conformal invariant,  despite that one of the solutions is NAF outside the scope of the theorem in \cite{jacobson}. However, some structures from the EMD theory do not perform the conformal invariance, as shown in \cite{kirill1,vanzo}.
\par
Through the more usual semi-classical methods for obtaining Hawking temperature, we get the proof of the conformal invariance of the temperature, in the subsections $5.1$-$5.4$. The result for a particular case of the conformal black hole in (\ref{34}), for $\gamma=0$, is that the temperature is a conformal invariant for all the methods. In the case of the Bogoliubov coefficients, the main feature of the conformal invariance, is given by the conformal invariance of the tortuous coordinate $ r^{*} $ and the near horizon limit, providing an invariant coordinate $t(T)$. For the metric euclidianization, the invariance of the tortuous coordinate $r^{*}$ is what characterizes the property of invariance of the temperature. In the case of the reflection coefficient, we found clearly a feature, but it seems that the limit of high frequencies $\omega_{1}$, of the quantized field, eliminates the dependence of the parameter $\omega$  of the conformal factor. Finally, the invariance obtained by the gravitational anomaly method appears to be linked to the invariance of the flow of energy-momentum tensor of the Hawking effect, as shown in \cite{jacobson}.
\par
We also have the so-called Hawking radiation by quantum tunneling approach \cite{ frank,rabin2}, which until now appeared to be an equivalent method to that of anomaly. In this approach, it seems that the conformal invariance result is maintained due to the similarity of obtaining the Hawking radiation with the anomaly method, being the temperature calculated essentially with the same expression (\ref{tha}).
\par
A work of Martinetti \cite{martinetti} analyzes the Unruh temperature through a conformal transformation. The result is that the Unruh temperature is not a conformal invariant. The Unruh temperature, or the Unruh effect, was used as an analogy to the Hawking temperature, or the Hawking effect \cite{davies2}. But as we saw, the Hawking temperature, in a specific case, is a conformal invariant, which is not the case of the Unruh temperature under the same conditions. This is due to the fact that the structure of a Rindler spacetime, regions R and L or R and L cones, is conformally transformed in another structure called double-cone. In these conditions, the analogy between the two quantum effects becomes incompatible.

\vspace{0,5cm}

{\bf Acknowledgement}: M. E. Rodrigues thanks M. J. S. Houndjo for the help in the elaboration and organization of the manuscript and the UFES for the hospitality during the development of this work and G.T.M. thanks CNPq and  FAPESPA for partial financial support.


\end{document}